\documentclass[aps,prd,showpacs,nofootinbib,twocolumn,widetext,amsmath,amssymb]{revtex4}
\usepackage{latexsym}
\usepackage{amsfonts}
\usepackage{amsbsy}
\usepackage{mathrsfs}

\begin{document}
\title{Gauge systems with noncommutative phase space}
\date{\today}

\author{Vladimir Cuesta}
\email{vcuesta@fis.cinvestav.mx} \affiliation{Departamento de F\'{\i}sica,
Cinvestav, Av. Instituto Polit\'ecnico Nacional 2508, San Pedro Zacatenco,
07360, Gustavo A. Madero, Ciudad de M\'exico, M\'exico}

\author{Merced Montesinos} \email{merced@fis.cinvestav.mx}
\affiliation{Departamento de F\'{\i}sica, Cinvestav, Av. Instituto
Polit\'ecnico Nacional 2508, San Pedro Zacatenco, 07360, Gustavo A. Madero,
Ciudad de M\'exico, M\'exico}

\author{Jos\'e David Vergara}
\email{vergara@nucleares.unam.mx} \affiliation{Instituto de Ciencias
Nucleares, Universidad Nacional Aut\'onoma de M\'exico, 70-543,
Ciudad de M\'exico, M\'exico}

\begin{abstract}
Some very simple models of gauge systems with noncanonical symplectic
structures having $sl(2,r)$ as the gauge algebra are given. The models can be
interpreted as noncommutative versions of the usual $SL(2,\mathbb{R})$ model
of Montesinos-Rovelli-Thiemann. The symplectic structures of the
noncommutative models, the first-class constraints, and the equations of
motion are those of the usual $SL(2,\mathbb{R})$ plus additional terms that
involve the parameters $\theta^{\mu\nu}$ which encode the noncommutativity
among the coordinates plus terms that involve the parameters $\Theta_{\mu\nu}$
associated with the noncommutativity among the momenta. Particularly
interesting is the fact that the new first-class constraints get corrections
linear and quadratic in the parameters $\theta^{\mu\nu}$ and
$\Theta_{\mu\nu}$. The current constructions show that noncommutativity of
coordinates and momenta can coexist with a gauge theory by explicitly building
models that encode these properties. This is the first time models of this
kind are reported which might be significant and interesting to the
noncommutative community.
\end{abstract}

\pacs{02.40.Gh, 11.10.Nx, 11.15.-q} \maketitle

Hamiltonian constrained systems with a finite number of degrees of freedom can
be written in a Hamiltonian form by means of the dynamical equations of motion
(summation convention over repeated indices is used throughout)
\begin{eqnarray}\label{demotion}
{\dot x}^{\mu} &=& \omega^{\mu\nu} (x) \left ( \frac{\partial
H}{\partial x^{\nu}} + \lambda^a \frac{\partial \gamma_a}{\partial
x^{\nu}} + \lambda^{\alpha} \frac{\partial \chi_{\alpha}}{\partial
x^{\nu}} \right )
\nonumber\\
&=& \omega^{\mu\nu}(x) \frac{\partial H_E}{\partial x^{\nu}}, \quad
\mu,\nu=1,2,\ldots,2N,
\end{eqnarray}
where $H_E = H+ \lambda^a \gamma_a + \lambda^{\alpha} \chi_{\alpha}$
is the extended Hamiltonian,
\begin{eqnarray}\label{const}
\gamma_a (x) \approx 0, \quad \chi_{\alpha} (x) \approx 0,
\end{eqnarray}
are the constraints which define the constraint surface $\Sigma$ embedded in
the phase space $\Gamma$. $\Gamma$ is a symplectic manifold endowed with the
symplectic structure
\begin{eqnarray}
\omega=\frac12 \omega_{\mu\nu} (x) d x^{\mu} \wedge d x^{\nu},
\end{eqnarray}
where $(x^\mu)$ are coordinates which locally label the points ${\bf p}$ of
$\Gamma$. It is important to emphasize that the phase space $\Gamma$ is
considered as a single entity, i.e., $\Gamma$ need not be necessarily
interpreted as the cotangent bundle of a configuration space ${\cal C}$. $H$
is taken to be a first-class Hamiltonian, the $\gamma$'s are first-class
constraints while the $\chi$'s are second class, i.e.,
\begin{subequations}
\label{alt}
\begin{equation}
\{ \gamma_a , \gamma_b \}_{\omega} = C_{ab}\,^c \gamma_c +
T_{ab}\,^{\alpha\beta} \chi_{\alpha} \chi_{\beta}, \label{1alt}
\end{equation}
\begin{equation}
\{ \gamma_a , \chi_{\alpha} \}_{\omega} = C_{a \alpha}\,^b \gamma_b
+ C_{a\alpha}\,^{\beta} \chi_{\beta},
\end{equation}
\begin{equation}
\{ H , \gamma_a \}_{\omega} = V_a\,^b \gamma_b + V_a\,^{\alpha\beta}
\chi_{\alpha} \chi_{\beta},
\end{equation}
\begin{equation}
 \{ H , \chi_{\alpha} \}_{\omega} =
V_{\alpha}\,^b \gamma_b + V_{\alpha}\,^{\beta} \chi_{\beta}.
\end{equation}
\end{subequations}

Also,
\begin{eqnarray}
\{ \chi_{\alpha}, \chi_{\beta} \}_{\omega} = C_{\alpha\beta}, \quad
\det{\left ( C_{\alpha\beta} \right )} \neq 0, \label{5alt}
\end{eqnarray}
where the Poisson brackets $\{\,\,,\}_{\omega}$ involved in Eqs. (\ref{alt})
and (\ref{5alt}) are computed using the symplectic structure $\omega$ on
$\Gamma$:
\begin{eqnarray}\label{pb}
\{ f , g \}_{\omega} = \frac{\partial f}{\partial x^{\mu}}
\omega^{\mu\nu} (x) \frac{\partial g}{\partial x^{\nu}}.
\end{eqnarray}
The dynamical equations of motion (\ref{demotion}) and the
constraints (\ref{const}) can be obtained from the action principle
\cite{MM2004,Mon2005}
\begin{eqnarray}\label{action} S [x^{\mu},
\lambda^a , \lambda^{\alpha} ] = \int^{\tau_2}_{\tau_1} \left ( \theta_{\mu}
(x) {\dot x}^{\mu} - H_E \right ) d \tau,
\end{eqnarray}
where $H_E= H + \lambda^a \gamma_a  + \lambda^{\alpha} \chi_{\alpha}$ as
before, $\omega = d \theta$ with $\theta= \theta_{\mu} (x) d x^{\mu}$ the
symplectic potential 1-form. It is important to emphasize that this way of
formulating Hamiltonian constrained systems is beyond Dirac's method which, by
construction, employs canonical symplectic structures \cite{D1,D2,D3,Dbook}.

In order to set down the problem addressed in the paper which links the use of
noncanonical symplectic structures in gauge theories to noncommutative
geometry, it is convenient to say some words about the various notions of
noncommutative geometry used by physicists.

1. First of all, there exists the so-called field theory defined on spacetimes
endowed with noncommuting coordinates \cite{snyder,ncg1}, where the
noncommutativity of the coordinates requires a noncommutativity among the
field variables. In this sense, most of the work in this field of research
consists in to mix consistently the noncommutativity of the coordinates of
spacetime with the noncommutativity among the field variables
\cite{chai,chaichian}.

2. On the other hand, there exists a version where the noncommutativity of the
coordinates of spacetime modifies the dynamics of test particles moving in it,
in the sense that the equations of motion for the test particles get
corrections that involve the parameters attached to the noncommuting
coordinates \cite{Li}. There, the situation usually studied involves a
Hamiltonian system without constraints, endowed with a canonical symplectic
structure $\Omega= d p_i \wedge d q^i$, or, equivalently,
\begin{eqnarray}
\{ q^i , q^j \}_{\Omega}= 0, \{q^i , p_j \}_{\Omega}= \delta^i_j, \{
p_i , p_j \}_{\Omega}=0,
\end{eqnarray}
and with an action principle of the form
\begin{eqnarray}\label{ncomm}
S[q^i , p_i] = \int^{t_2}_{t_1} \left ( p_i {\dot q}^i - H (q,p,t)
\right ) dt,
\end{eqnarray}
as a starting point. Next, the dynamical equations obtained from the action of
Eq. (\ref{ncomm}) are {\it ad hoc} modified by replacing the original
symplectic structure $\Omega$ with a new one $\omega_{\mbox{nc}}$ which
involves noncommutativity among the coordinates $q$'s and among the $p$'s
\begin{equation}
\{ q^i , q^j \}_{\omega_{\mbox{\tiny nc}}} = \theta^{ij},\; \{ q^i , p_j
\}_{\omega_{\mbox{\tiny nc}}} = \delta^i_j, \; \{ p_i , p_j
\}_{\omega_{\mbox{\tiny nc}}} = \Theta_{ij}.
\end{equation}
$\theta^{ij}=-\theta^{ji}$ and $\Theta_{ij}=-\Theta_{ji}$ are, usually,
assumed to be constant parameters. The conceptual (or physical) explanation
about the origin of noncommutativity depends on the particular theory assumed
to be the fundamental one. Independently of the explanation, some consequences
of the introduction of $\omega_{\mbox{\tiny nc}}$ are: (1) the original
symmetries of the various geometrical objects involved might change, in
particular, the algebra of observables ${\cal A}_{\omega_{\mbox{\tiny nc}}}$
(computed using $\omega_{\mbox{\tiny nc}}$) might be {\it completely
different} to the original one ${\cal A}_{\Omega}$ (computed using $\Omega$),
(2) the elements that form a complete set of commuting observables (csco)
which are in involution with respect to $\Omega$ might be {\it completely
different} to the elements that form a complete set of commuting observables
computed with respect to $\omega_{\mbox{\tiny nc}}$, in fact the elements of
the former set are not in involution with respect to $\omega_{\mbox{\tiny
nc}}$ in the generic case.

Therefore, it is rather natural to ask if there exists analogs for
gauge systems of the situation previously mentioned in the item 2.
In this paper, it is shown that the answer is in the affirmative and
the term ``gauge systems with noncommutative phase space" is used to
refer to this fact, that is to say, that it is technically possible
to start from a gauge system endowed with a canonical symplectic
structure $\Omega$ and then to replace the original symplectic
structure $\Omega$ in $\Gamma$ with a new one $\omega_{\mbox{\small
nc}}$ to build a new gauge system in such a way that the resulting
gauge group (more precisely, the gauge algebra of the first-class
constraints) be the same than that of the original gauge system.
Even though the analysis will be focused on the item 2, it will
result clear at the end of the paper, that the current approach can
also be used in the context of item 1.

The current ideas will be implemented in a nontrivial gauge system, the
$SL(2,\mathbb{R})$ model. The phase space $\Gamma= \mathbb{R}^8$, whose points
are locally labeled by $(x^{\mu})=(u^1,u^2,v^1,v^2,p_1,p_2,\pi_1,\pi_2)$, is
endowed with the symplectic structure
\begin{eqnarray}\label{csesl2r}
\Omega = d p_i \wedge d u^i + d \pi_i \wedge d v^i, \quad i=1,2,
\end{eqnarray}
which has the canonical form, in total agreement with Dirac's method. The
model is defined by the action principle \cite{mrt}
\begin{eqnarray}\label{model}
& & S[u^i, v^i,p_i,\pi_i, N, M, \lambda] = \int^{\tau_2}_{\tau_1}
\left (
{\dot u}^i p_i + {\dot v}^i \pi_i \right. \nonumber\\
&& \mbox{} \left. - N H_1 - N H_2 - \lambda D \right ) d \tau,
\quad\quad i=1,2,
\end{eqnarray}
where the constraints
\begin{eqnarray}\label{fccsl2r}
H_1 &:=& \frac{1}{2} \left [ (p_1)^2 + (p_2)^2 - (v^1)^2 - (v^2)^2
\right ] \approx 0 ,\nonumber\\
H_2 &:=& \frac{1}{2} \left [ (\pi_1)^2 + (\pi_2)^2 - (u^1)^2 -
(u^2)^2 \right]
\approx 0 , \nonumber\\
D &:=& u^i p_i - v^i \pi_i \approx 0, \quad i=1,2,
\end{eqnarray}
satisfy the relations
\begin{eqnarray}\label{al}
\{ H_1 , H_2 \}_{\Omega} &=& D, \nonumber\\
\{ H_1 , D \}_{\Omega} &=& - 2 H_1, \nonumber\\
\{ H_2 , D \}_{\Omega} &=& 2 H_2.
\end{eqnarray}
which turn out to be isomorphic to those defining the $sl(2,r)$ Lie algebra.

Before going on, it is convenient to say that in order to make clear the
concepts involved in the paper, the original $SL(2,\mathbb{R})$ model will be
gradually modified, by first introducing one parameter, next two parameters,
and finally, the generic case which allows noncommutativity among the
coordinates and among the momenta involved. It will be shown for the case of
the $SL(2,\mathbb{R})$ model that it is always possible, at least locally, to
maintain the same algebra and obtain as a consequence a new set of constraints
that corresponds to the twisted generators of the original symmetry
\cite{chai}.

{\it Model with one noncommutative parameter $\theta$}. As it was mentioned,
the idea behind the current theoretical framework is to modify the symplectic
structure (\ref{csesl2r}) of the original $SL(2,\mathbb{R})$ model by
introducing a parameter $\theta$ that encodes the noncommutativity among the
coordinates, the phase space of the system is still $\Gamma=\mathbb{R}^8$, its
points are still locally labeled by
$(x^{\mu})=(u^1,u^2,v^1,v^2,p_1,p_2,\pi_1,\pi_2)$, but now the inverse of the
new symplectic structure on it is given by
\begin{eqnarray}\label{sedavid}
\left ( \omega^{\mu\nu}_{\mbox{\tiny nc}} \right ) &=& \left (
\begin{array}{rrrrrrrr}
0 & \theta & 0 & 0 & 1 & 0 & 0 & 0 \\
- \theta & 0 & 0 & 0 & 0 & 1 & 0 & 0 \\
0 & 0 & 0 & 0 & 0 & 0 & 1 & 0 \\
0 & 0 & 0 & 0 & 0 & 0 & 0 & 1 \\
-1 & 0 & 0 & 0 & 0 & 0 & 0 & 0 \\
0 & -1 & 0 & 0 & 0 & 0 & 0 & 0 \\
0 & 0 & -1 & 0 & 0 & 0 & 0 & 0 \\
0 & 0 & 0 & -1 & 0 & 0 & 0 & 0
\end{array}
\right ).
\end{eqnarray}
Nevertheless, the original constraints (\ref{fccsl2r}) do {\it not} close
under the bracket $\{A,B\}_{\omega_{\mbox{\tiny nc}}}$ computed with
(\ref{sedavid}). In order to preserve the $SL(2,\mathbb{R})$ symmetry of the
original system in the new model, it is mandatory to modify the original
constraints. The more efficient way to build the new first class constraints
is by means of Darboux's theorem \cite{Arnold}. According to it, it is always
possible, at least locally, to find a map from the original coordinate system
and symplectic structure (\ref{sedavid}) to a new coordinate system (Darboux's
variables) in terms of which the symplectic structure acquires the ordinary
canonical form. In the current case this map is given by
\begin{eqnarray}
\tilde u^1 &=& u^1 + \theta p_2 \nonumber\\
\tilde u^2 &=& u^2,\dots
\end{eqnarray}
Now, with respect to Darboux's variables $(\tilde u^1,\tilde u^2,\tilde
v^1,\tilde v^2,\tilde p_1,\tilde p_2,\tilde \pi_1,\tilde \pi_2)$, the new
constraints expressed in terms of the new variables have the same analytical
form than the old constraints have in terms of the old variables. In terms of
the original variables, the new constraints acquire the form
\begin{eqnarray}\label{ncc}
{\cal H}_1 &:=& \frac12 \left [ (p_1)^2 + (p_2)^2 - (v^1)^2 -
(v^2)^2 \right ]
\approx 0, \nonumber\\
{\cal H}_2 &:=& \frac12 \left [ (\pi_1)^2 + (\pi_2)^2 - (u^1)^2 -
(u^2)^2
\right ] \nonumber\\
& & \mbox{} - \theta u^1 p_2 - \frac12 \theta^2 \left ( p_2 \right
)^2 \approx 0,
\nonumber\\
{\cal D} &:=& u^i p_i - v^i \pi_i + \theta p_1 p_2 \approx 0, \quad
i=1,2.
\end{eqnarray}
It turns out that the constraints (\ref{ncc}) are also first-class, but {\it
now} with respect to the Poisson brackets computed with the new symplectic
structure of Eq. (\ref{sedavid}). The resulting algebra of constraints is
\begin{eqnarray}\label{ncal}
\{ {\cal H}_1 , {\cal H}_2 \}_{\omega_{\mbox{\tiny nc}}} &=& {\cal D}, \nonumber\\
\{ {\cal H}_1 , {\cal D} \}_{\omega_{\mbox{\tiny nc}}} &=& - 2 {\cal H}_1, \nonumber\\
\{ {\cal H}_2 , {\cal D} \}_{\omega_{\mbox{\tiny nc}}} &=& 2 {\cal H}_2,
\end{eqnarray}
which is isomorphic to the $sl(2,r)$ Lie algebra, as expected. By inserting
(\ref{sedavid}) and (\ref{ncc}) with
$(\lambda^1,\lambda^2,\lambda^3)=(N,M,\lambda)$ and $(\gamma_1,
\gamma_2,\gamma_3)=({\cal H}_1, {\cal H}_2, {\cal D})$ into Eqs.
(\ref{demotion}) leads to
\begin{eqnarray}\label{deford}
{\dot u}^1 &=& N p_1 - M \theta u^2 + \lambda \left ( u^1 + 2 \theta p_2 \right ), \nonumber\\
{\dot u}^2 &=& N p_2 + \lambda u^2, \nonumber\\
{\dot v}^1 &=& M \pi_1 - \lambda v^1, \nonumber\\
{\dot v}^2 &=& M \pi_2 - \lambda v^2, \nonumber\\
{\dot p}_1 &=& M \left ( u^1 + \theta p_2 \right ) - \lambda p_1, \nonumber\\
{\dot p}_2 &=& M u^2 - \lambda p_2, \nonumber\\
{\dot \pi}_1 &=& N v^1 + \lambda \pi_1, \nonumber\\
{\dot \pi}_2 &=& N v^2 + \lambda \pi_2,
\end{eqnarray}
which are {\it different} from those corresponding to the original
$SL(2,\mathbb{R})$ system because there are terms that involve
$\theta$. Note that when $\theta=0$, the symplectic structure
(\ref{sedavid}), the constraints (\ref{ncc}), and the dynamical
equations of motion (\ref{deford}) acquire the same form of those
corresponding to the original $SL(2,\mathbb{R})$ model.

Using (\ref{deford}), the evolution of the constraints (\ref{ncc})
yields
\begin{eqnarray}
{\dot {\cal H}_1} &=& M {\cal D} - 2 \lambda {\cal H}_1, \nonumber\\
{\dot {\cal H}_2} &=& - N {\cal D} + 2 \lambda {\cal H}_2, \nonumber\\
{\dot {\cal D}} &=& - 2M {\cal H}_2 + 2 N {\cal H}_1,
\end{eqnarray}
in agreement with the $sl(2,r)$ Lie algebra (\ref{ncal}).

Observables. It can be shown that the following six observables
\begin{eqnarray}
{\cal O}_{12} &=& u^1 p_2 - p_1 u^2 + \theta (p_2)^2, \nonumber\\
{\cal O}_{13} &=& u^1 v^1 - p^1 \pi^1 + \theta v^1 p_2, \nonumber\\
{\cal O}_{14} &=& u^1 v^2 - p_1 \pi_2 + \theta v^2 p_2, \nonumber\\
{\cal O}_{23} &=& u^2 v^1 - p_2 \pi_1, \nonumber\\
{\cal O}_{24} &=& u^2 v^2 - p_2 \pi_2, \nonumber\\
{\cal O}_{34} &=& \pi_1 v^2 - v^1 \pi_2,
\end{eqnarray}
are all gauge invariant. The resulting Lie algebra of their Poisson brackets
computed with respect to the symplectic structure (\ref{sedavid}) is
isomorphic to the Lie algebra of the Lie group $SO(2,2)$, i.e., the algebra of
observables is exactly the same than that of the original $SL(2,\mathbb{R})$
model. The four relations among the observables are the same than those of the
original SL(2,R) model but now the expression of the observables are
\begin{eqnarray}
\epsilon &=& \frac{u^1 p_2 - p_1 u^2 + \theta \left ( p_2 \right )^2}
{\mid u^1 p_2 - p_1 u^2 + \theta \left ( p_2 \right )^2 \mid}, \\
\epsilon' &=& \frac{\pi_1 v^2 - v^1 \pi_2 }
{\mid \pi_1 v^2 - v^1 \pi_2   \mid },\\
J &=& \mid u^1 p_2 - p_1 u^2 + \theta \left ( p_2 \right )^2 \mid, \\
\alpha &=& \arctan{\left ( \frac{u^1 v^2 - p^1 \pi^2 + \theta v^2 p_2}{u^1 v^1
- p_1 \pi_1 + \theta p_2 v^1} \right )}.
\end{eqnarray}

{\it Model with two noncommutative parameters $\theta$ and $\phi$}. Now, the
symplectic structure (\ref{csesl2r}) of the original $SL(2,\mathbb{R})$ model
is modified by introducing two parameters $\theta$ and $\phi$ which encode the
noncommutativity among the coordinates, the phase space of the system is still
$\Gamma=\mathbb{R}^8$, its points are still locally labeled by
$(x^{\mu})=(u^1,u^2,v^1,v^2,p_1,p_2,\pi_1,\pi_2)$, but now the inverse of the
new symplectic structure on it is given by
\begin{eqnarray}\label{vladse}
\left ( \omega^{\mu\nu}_{\mbox{\tiny nc}} \right ) &=& \left (
\begin{array}{rrrrrrrr}
0 & \theta & 0 & 0 & 1 & 0 & 0 & 0 \\
- \theta & 0 & 0 & 0 & 0 & 1 & 0 & 0 \\
0 & 0 & 0 & \phi & 0 & 0 & 1 & 0 \\
0 & 0 & - \phi & 0 & 0 & 0 & 0 & 1 \\
-1 & 0 & 0 & 0 & 0 & 0 & 0 & 0 \\
0 & -1 & 0 & 0 & 0 & 0 & 0 & 0 \\
0 & 0 & -1 & 0 & 0 & 0 & 0 & 0 \\
0 & 0 & 0 & -1 & 0 & 0 & 0 & 0
\end{array}
\right ).
\end{eqnarray}
As for the case of the model with one parameter $\theta$, the
original constraints of Eq. (\ref{fccsl2r}) get also modified, by
terms linear and quadratic in the noncommutative parameters $\theta$
and $\phi$. The new constraints are
\begin{eqnarray}\label{ncc2}
{\cal C}_1 &:=& \frac12 \left [ (p_1)^2 + (p_2)^2 - (v^1)^2 -
(v^2)^2 \right ]
\nonumber\\
&& \mbox{} - \phi v^1 \pi_2 - \frac12 \phi^2 \left ( \pi_2 \right
)^2
\approx 0, \nonumber\\
{\cal C}_2 &:=& \frac12 \left [ (\pi_1)^2 + (\pi_2)^2 - (u^1)^2 -
(u^2)^2
\right ] \nonumber\\
&& \mbox{} - \theta u^1 p_2 - \frac12 \theta^2 \left ( p_2 \right
)^2 \approx
0, \nonumber\\
{\cal V} &:=& u^i p_i - v^i \pi_i + \theta p_1 p_2 - \phi \pi_1
\pi_2 \approx 0,
\end{eqnarray}
with $i=1,2$ [cf. with Eqs. (\ref{ncc})]. It turns out that the
constraints of Eq. (\ref{ncc2}) are also first-class but {\it now}
with respect to the Poisson brackets computed with the new
symplectic structure of Eq. (\ref{vladse}). The resulting algebra of
constraints is
\begin{eqnarray}\label{ncal2}
\{ {\cal C}_1 , {\cal H}_2 \}_{\omega_{\mbox{\tiny nc}}} &=& {\cal V}, \nonumber\\
\{ {\cal C}_1 , {\cal V} \}_{\omega_{\mbox{\tiny nc}}} &=& - 2 {\cal C}_1, \nonumber\\
\{ {\cal C}_2 , {\cal V} \}_{\omega_{\mbox{\tiny nc}}} &=& 2 {\cal C}_2,
\end{eqnarray}
which is isomorphic to the $sl(2,r)$ Lie algebra.

By plugging (\ref{vladse}), (\ref{ncc2}),
$(\lambda^1,\lambda^2,\lambda^3)=(N,M,\lambda)$, and $(\gamma_1,
\gamma_2,\gamma_3)=({\cal C}_1, {\cal C}_2, {\cal V})$ into Eqs.
(\ref{demotion}), the dynamical equations (\ref{demotion}) acquire the form
\begin{eqnarray}\label{deford2}
{\dot u}^1 &=& N p_1 - M \theta u^2 + \lambda \left ( u^1 + 2 \theta p_2 \right ), \nonumber\\
{\dot u}^2 &=& N p_2 + \lambda u^2, \nonumber\\
{\dot v}^1 &=& M \pi_1 - \lambda \left ( v^1 + 2 \phi \pi_2 \right ) - \phi N v^2, \nonumber\\
{\dot v}^2 &=& M \pi_2 - \lambda v^2, \nonumber\\
{\dot p}_1 &=& M \left ( u^1 + \theta p_2 \right ) - \lambda p_1, \nonumber\\
{\dot p}_2 &=& M u^2 - \lambda p_2, \nonumber\\
{\dot \pi}_1 &=& N \left ( v^1 + \phi \pi_2 \right ) + \lambda \pi_1, \nonumber\\
{\dot \pi}_2 &=& N v^2 + \lambda \pi_2.
\end{eqnarray}
[cf. with Eqs. (\ref{deford})]. Using (\ref{deford2}), the evolution
of the constraints (\ref{ncc2}) yields
\begin{eqnarray}
{\dot {\cal C}_1} &=& M {\cal V} - 2 \lambda {\cal C}_1, \nonumber\\
{\dot {\cal C}_2} &=& - N {\cal V} + 2 \lambda {\cal C}_2, \nonumber\\
{\dot {\cal V}} &=& - 2M {\cal C}_2 + 2 N {\cal C}_1,
\end{eqnarray}
in agreement with the $sl(2,r)$ Lie algebra (\ref{ncal2}).

Observables. The following six observables
\begin{eqnarray}
{\cal O}_{12} &=& u^1 p_2 - p_1 u^2 + \theta \left ( p_2 \right )^2, \nonumber\\
{\cal O}_{13} &=& u^1 v^1 - p^1 \pi^1 + \theta v^1 p_2
+ \theta \phi p_2 \pi_2 + \phi u^1 \pi_2 , \nonumber\\
{\cal O}_{14} &=& u^1 v^2 - p_1 \pi_2 + \theta v^2 p_2, \nonumber\\
{\cal O}_{23} &=& u^2 v^1 - p_2 \pi_1 + \phi u^2 \pi_2, \nonumber\\
{\cal O}_{24} &=& u^2 v^2 - p_2 \pi_2, \nonumber\\
{\cal O}_{34} &=& \pi_1 v^2 - v^1 \pi_2 - \phi \left ( \pi_2 \right
)^2,
\end{eqnarray}
are all gauge invariant. Again, the resulting Lie algebra of their Poisson
brackets computed with respect to the symplectic structure (\ref{vladse}) is
isomorphic to the Lie algebra of the Lie group $SO(2,2)$, i.e., the algebra of
observables is exactly the same than that of the original $SL(2,\mathbb{R})$
model. The four relations among the observables is the same than those of the
original SL(2,R) model but now the expression of the observables are
\begin{eqnarray}
\epsilon &=& \frac{u^1 p_2 - p_1 u^2 + \theta \left ( p_2 \right )^2}
{\mid u^1 p_2 - p_1 u^2 + \theta \left ( p_2 \right )^2 \mid}, \\
\epsilon' &=& \frac{\pi_1 v^2 - v^1 \pi_2 - \phi \left ( \pi_2 \right )^2}
{\mid \pi_1 v^2 - v^1 \pi_2 - \phi \left ( \pi_2 \right )^2  \mid },\\
J &=& \mid u^1 p_2 - p_1 u^2 + \theta \left ( p_2 \right )^2 \mid, \\
\alpha &=& \arctan{\left ( \frac{u^1 v^2 - p^1 \pi^2 + \theta v^2 p_2}{u^1 v^1
- p_1 \pi_1 + \theta p_2 v^1} \right )}.
\end{eqnarray}

{\it Model with noncommuting momenta and noncommuting coordinates}. The more
general case corresponds to consider noncommutativity among all the momenta
$p_\mu=(p_1,p_2,\pi_1,\pi_2)$ and among all the coordinates
$x^\mu=(u^1,u^2,v^1,v^2)$. In this case the noncommutative brackets are given
by:
\begin{eqnarray} \label{gral}
\{x^\mu,x^\nu \}_{\omega_{\mbox{\tiny nc}}} &=& \theta^{\mu\nu}, \nonumber \\
\{p_\mu,p_\nu \}_{\omega_{\mbox{\tiny nc}}} &=& \Theta_{\mu\nu}, \\
\{x^\mu,p_\nu \}_{\omega_{\mbox{\tiny nc}}} &=& \delta^\mu_{\ \nu}, \nonumber
\end{eqnarray}
where, in order that could be possible to get a representation in terms of
commutative variables \cite{Li}, the noncommutative parameters are restricted
by $\theta^{\mu\nu} \Theta_{\nu\rho}= 4\alpha\beta(\alpha\beta
-1)\delta^\mu_{\ \rho}$, with $(\alpha,\beta)$ arbitrary real parameters.
Taking this relation into account the commutative variables are
\begin{eqnarray}
\tilde x^\mu &=& \frac{\beta}{\rho}x^\mu + \frac{\theta^{\mu\nu}}{2\alpha\rho}
p_\nu,\nonumber\\
\tilde p_\mu &=&\frac{\alpha}{\rho}p_\mu
-\frac{\Theta_{\mu\nu}}{2\beta\rho}x^\nu,
\end{eqnarray}
with $\rho= 2\alpha\beta -1 $ and the new constraints are
\begin{eqnarray}\label{ncc3}
\gamma_1 &:=& \frac12 \left [ (\tilde p_1)^2 + (\tilde p_2)^2 - (\tilde v^1)^2
- (\tilde v^2)^2 \right ]
\approx 0, \nonumber\\
\gamma_2 &:=& \frac12 \left [ (\tilde \pi_1)^2 + (\tilde \pi_2)^2 - (\tilde
u^1)^2 - (\tilde u^2)^2 \right ] \approx
0, \nonumber\\
d &:=& \tilde u^i \tilde p_i - \tilde v^i \tilde \pi_i \approx 0.
\end{eqnarray}
The algebra of these constraints will be isomorphic to the $sl(2,r)$ Lie
algebra by construction. However, when these constraints are written
explicitly in terms of the noncommuting variables (\ref{gral}), they become
the twist generators of the $sl(2,r)$ algebra. Also, for this noncommutative
model, it is possible to construct Dirac's observables
\begin{eqnarray}
{\cal O}_{12} &=& \tilde u^1 \tilde p_2 - \tilde p_1 \tilde u^2, \nonumber\\
{\cal O}_{13} &=& \tilde u^1 \tilde v^1 - \tilde p^1 \tilde \pi^1, \nonumber\\
{\cal O}_{14} &=& \tilde u^1 \tilde v^2 - \tilde p_1 \tilde \pi_2, \nonumber\\
{\cal O}_{23} &=& \tilde u^2 \tilde v^1 - \tilde p_2 \tilde \pi_1, \nonumber\\
{\cal O}_{24} &=& u^2 v^2 - p_2 \pi_2, \nonumber\\
{\cal O}_{34} &=& \tilde \pi_1 \tilde v^2 -\tilde v^1 \tilde \pi_2,
\end{eqnarray}
which are all gauge invariant under the transformation generated by the first
class constraints.

Furthermore, the resulting Lie algebra of their Poisson brackets computed with
respect to the symplectic structure (\ref{gral}) is isomorphic to the Lie
algebra of the Lie group $SO(2,2)$, i.e., the algebra of observables is
exactly the same than that of the original $SL(2,\mathbb{R})$ model.

The relevance of the noncommutative models of this paper lies in the
fact that it has been explicitly shown that it possible to introduce
noncommutativity among coordinates and among momenta which are both
compatible with the gauge principle in the sense that algebra of
generators closes [cf. with Ref. \cite{ture}]. The result shows that
it is possible, in the context of noncommutative quantum mechanics,
to twist the gauge symmetry in agreement with a recent proposal in
field theory\cite{wess}. The generalization of these results to
noncommutative field theory (item 1 of this paper) would mean that
it is possible to modify the usual symplectic structure among the
field variables by symplectic structures involving noncommuting
field variables. Due to the fact that the noncommutativity among the
gauge fields is related to the noncommutativity of the coordinates
of spacetime, a mechanism to combine consistently the two types of
noncommutativity must be required, for instance incorporating the
coordinates as dynamical variables (e.g., parameterizing the field
theory) to set noncommuting coordinates of spacetime and
noncommuting fields on the same footing \cite{rosver}. By doing
things in this way, both types of noncommutativity would emerge in
an extended phase space endowed with noncommutative symplectic
structures, situation that can be handled with the same ideas
developed in this work.

\section*{Acknowledgements}
This work was supported in part by CONACyT Grant Nos. SEP-2003-C02-43939 and
47211. J.D. Vergara also acknowledges support from DGAPA-UNAM Grant No.
IN104503.


\end{document}